\documentclass[final]{siamltex}
\usepackage{graphicx,amsfonts,amsmath,epsfig,cite,url}

\begin{document}

\setcounter{page}{0}
\thispagestyle{empty}
\begin{center}
{\huge Communities in Networks \\} 
\vspace{.2 in}
{\large Mason A. Porter, Jukka-Pekka Onnela, and Peter J. Mucha}
\end{center}

\begin{figure}[h]
\centerline{
\includegraphics[width = .8\textwidth]{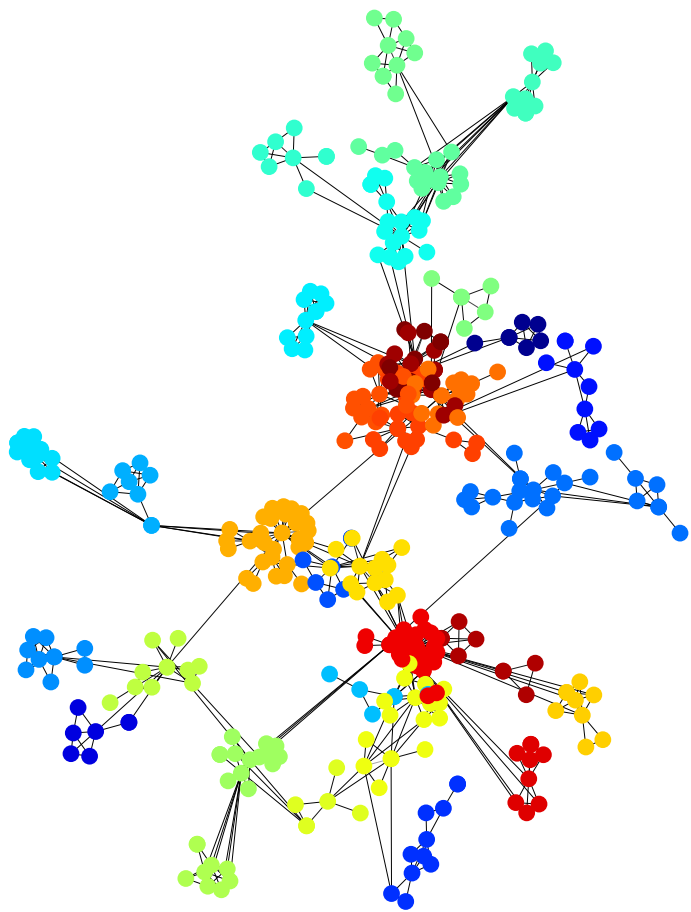}
}
\caption{[{Proposed cover figure.}] The largest connected component of a network of network scientists.  This network was constructed based on the coauthorship of papers listed in two well-known review articles \cite{newmansirev,bocca06} and a small number of additional papers that were added manually \cite{newmodlong}.  Each node is colored according to community membership, which was determined using a leading-eigenvector spectral method followed by Kernighan-Lin node-swapping steps \cite{newmodlong,kl,richardson}.  To determine community placement, we used the Fruchterman-Reingold graph visualization \cite{fr91}, a force-directed layout method that is related to maximizing a quality function known as modularity \cite{noack08}.  To apply this method, we treated the communities as if they were themselves the nodes of a (significantly smaller) network with connections rescaled by inter-community links.  We then used the Kamada-Kawaii spring-embedding graph visualization algorithm \cite{kk} to place the nodes of each individual community (ignoring inter-community links) and then to rotate and flip the communities for optimal placement (including inter-community links).  We gratefully acknowledge Amanda Traud for preparing this figure.}
\label{cover}
\end{figure}

\clearpage


\title{Communities in Networks}

\author{Mason A. Porter$^{1,4}$, Jukka-Pekka Onnela$^{2,3,4,5}$, and Peter J. Mucha$^{6,7}$
  \\
  $^1$\footnotesize{Oxford Centre for Industrial and Applied Mathematics, Mathematical Institute, University of Oxford, OX1 3LB, UK} \\
  $^2$\footnotesize{Harvard Kennedy School, Harvard University, Cambridge, MA  02138, USA} \\  
    $^3$\footnotesize{Department of Physics, University of Oxford, OX1 1HP, UK} \\
$^4$\footnotesize{CABDyN Complexity Centre, University of Oxford, OX1 1HP, UK} \\
$^5$\footnotesize{Department of Biomedical Engineering and Computational Science, Helsinki University of Technology, P.O. Box 9203, FI-02015 TKK, Finland } \\
  $^6$\footnotesize{Carolina Center for Interdisciplinary Applied Mathematics, Department of Mathematics, University of North Carolina,\\ Chapel Hill, NC 27599-3250, USA} \\
  $^7$\footnotesize{Institute for Advanced Materials, Nanoscience and Technology, University of
    North Carolina, Chapel Hill, NC 27599, USA} }

\maketitle

\vspace{.2 in}








\section*{Introduction: Networks and Communities} 

\textit{``But although, as a matter of history, statistical mechanics owes its origin to investigations in
thermodynamics, it seems eminently worthy of an independent development, both on account of the elegance and simplicity of its principles, and because it yields new results and places old truths in
a new light in departments quite outside of thermodynamics.''}

\vspace{.05 in}

-- Josiah Willard Gibbs, \textit{Elementary Principles in Statistical Mechanics}, 1902 \cite{gibbs}

\vspace{.15 in}

From an abstract perspective, the term \textit{network} is used as a
synonym for a mathematical \textit{graph}.  However, to scientists
across a variety of fields, this label means so much more
\cite{faust,freemanbook,str01,newmansirev,bocca06, cald,newmanphystoday}. In
sociology, each \textit{node} (or vertex) of a network represents an
\textit{agent}, and a pair of nodes can be connected by a
\textit{link} (or edge) that signifies some social
interaction or tie between them.  Each node has a \textit{degree} given by the number of edges connected to it and a \textit{strength} given by the total weight of those edges.
Graphs can represent either man-made or natural
constructs, such as the World Wide Web or
neuronal synaptic networks in the brain.  Agents in such
networked systems play the role of the particles in traditional statistical
mechanics that we all know and (presumably) love, and the structure 
of interactions between agents reflects the microscopic rules that govern
their behavior. The simplest types of links are binary
pairwise connections, in which one only cares about the presence or absence of a tie.
However, in many situations, links can also be assigned a direction and a
(positive or negative) weight to designate different interaction
strengths.

\begin{figure}[htbp]
\centerline{
{\epsfig{file=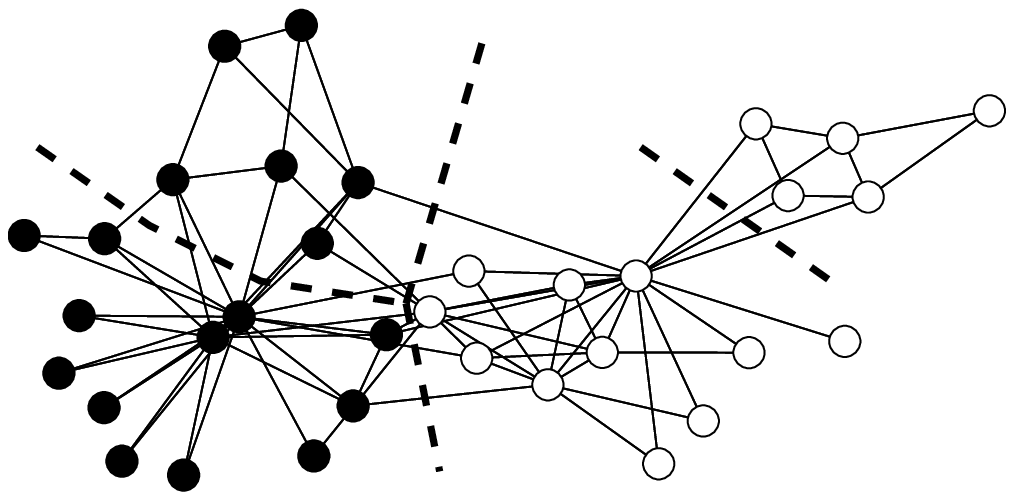, width=.62\textwidth}}
{\epsfig{file=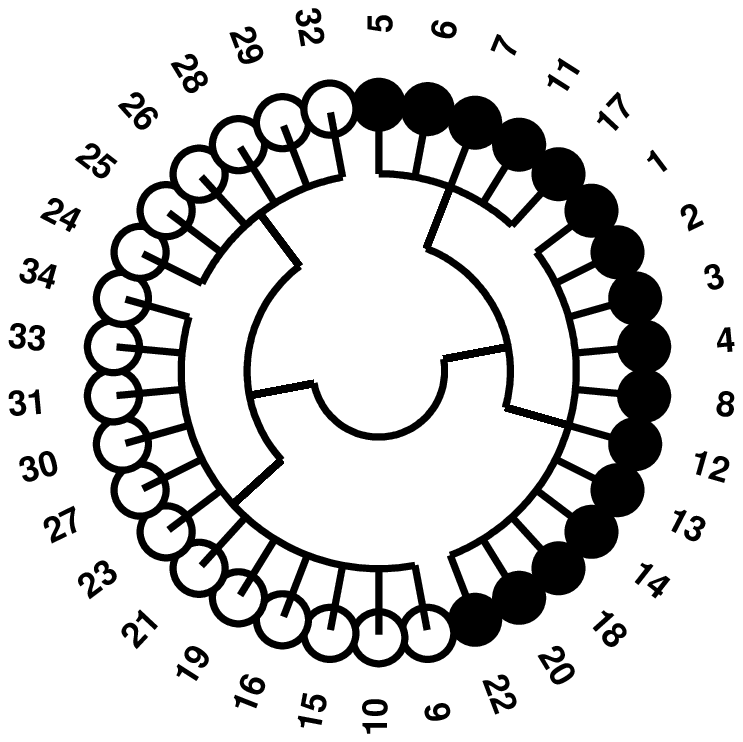, width=.38\textwidth}}
}
\caption{(Left) The Zachary karate club network \cite{karate}, visualized using the Fruchterman-Reingold method \cite{fr91}. Nodes are colored black or white depending on their later club affiliation (after a disagreement prompted the organization's breakup).  The dashed lines separate different communities, which were determined using a leading-eigenvector spectral maximization of modularity
\cite{newmodlong} with subsequent Kernighan-Lin node-swapping steps 
(see the discussion in the main text).  (Right) Polar coordinate dendrogram representing the results of applying this community-detection algorithm to the network.  Nodes are grouped into the communities indicated in the left panel.  One can see the initial split of the network into two branches (identical to the observed membership of the new clubs) by moving outward from the center of the ring.  Moving further outward culminates in the final partition of the network into four communities.}
\label{karateplot}
\end{figure}

Traditional statistical physics is concerned with the dynamics of
ensembles of interacting and non-interacting particles.  Rather than
tracking the motion of all of the particles simultaneously, which is
an impossible task due to their tremendous number, one averages (in
some appropriate manner) the microscopic rules that govern the dynamics of individual particles to make precise statements of macroscopic observables such as temperature and density \cite{schwabl}.  It is also sometimes possible to make comments about intermediate \textit{mesoscopic} structures, which lie between the microscopic and macroscopic worlds; they are large enough that it is reasonable to discuss their collective properties but small enough so that those properties are obtained through averaging over smaller numbers of constituent items.  One can similarly take a collection of interacting
agents, such as the nodes of a network, with some set of microscopic
interaction rules and attempt to derive the resulting mesoscopic and
macroscopic structures. One mesoscopic structure, called a \textit{community}, consists of a group
of nodes that are relatively densely connected to each other but 
sparsely connected to other dense groups
in the network \cite{santolong}.  We illustrate this idea in Fig.~\ref{karateplot} using a well-known benchmark network from the sociology literature \cite{karate}.  

The existence of social communities is intuitively clear, and the grouping patterns of humans have been studied for a long time in both sociology \cite{freemanbook,coleman1964,moody03} and social anthropology \cite{kottak1991,scupin1992}. Stuart Rice clustered data by hand to investigate political blocs in the 1920s \cite{rice1927}, and George Homans illustrated the usefulness of rearranging the rows and columns of data matrices to reveal their underlying structure in 1950 \cite{homans}. Robert Weiss and Eugene Jacobson performed (using organizational data) what may have been the first analysis of network community structure in 1955 \cite{weiss55}, and Herbert Simon espoused surprisingly modern views on community structure and complex systems in general in the 1960s \cite{herbert}. Social communities are indeed ubiquitous, arising in the flocking of animals and in social organizations in every type of human society: groups of hunter-gatherers, feudal structures, royal families, political and business organizations, families, villages, cities, states, nations, continents, and even virtual communities such as Facebook groups \cite{santolong,newmanphystoday}.  Indeed, the concept of community is one of everyday familiarity.  We are all connected to relatives, friends, colleagues, and acquaintances who are in turn connected to each other in groups of different sizes and cohesions.  The goals of studying social communities have aligned unknowingly with the statistical physics paradigm.  As sociologist Mark Granovetter wrote in his seminal 1973 paper \cite{weak} on weak ties, ``Large-scale statistical, as well as qualitative, studies offer a good deal of insight into such
macro phenomena as social mobility, community organization, and
political structure... But how interaction in small groups
aggregates to form large-scale patterns eludes us in most cases.''

\begin{figure}
\centerline{
\includegraphics[width = \textwidth]{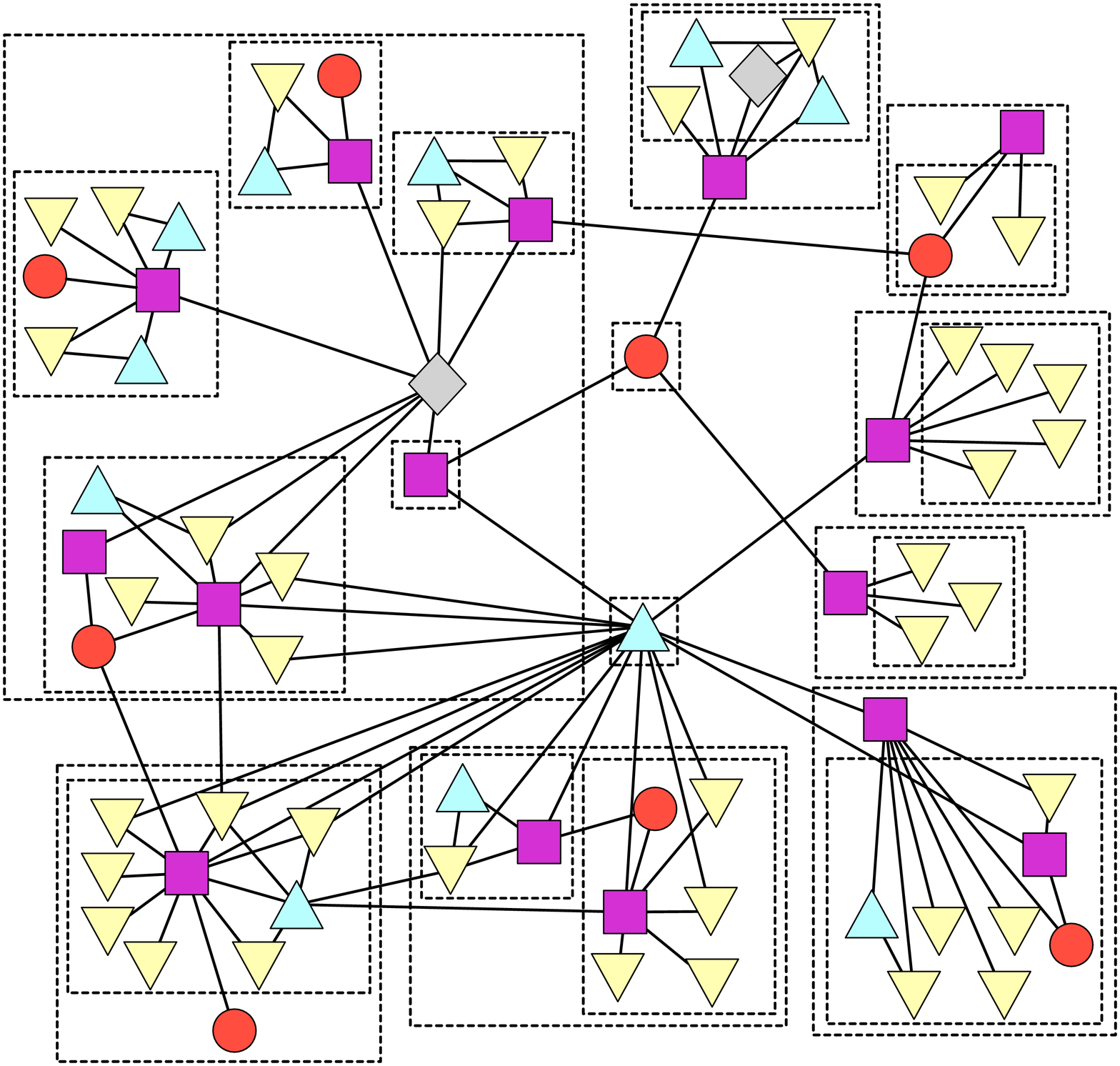}
}
\caption{A network with both hierarchical and modular structure.  This image, courtesy of Aaron Clauset, is an adaptation of a figure from Ref.~\cite{clausetnature}.}
\label{clausetfigure}
\end{figure}

\begin{figure}
\centerline{
\includegraphics[width = .5\textwidth]{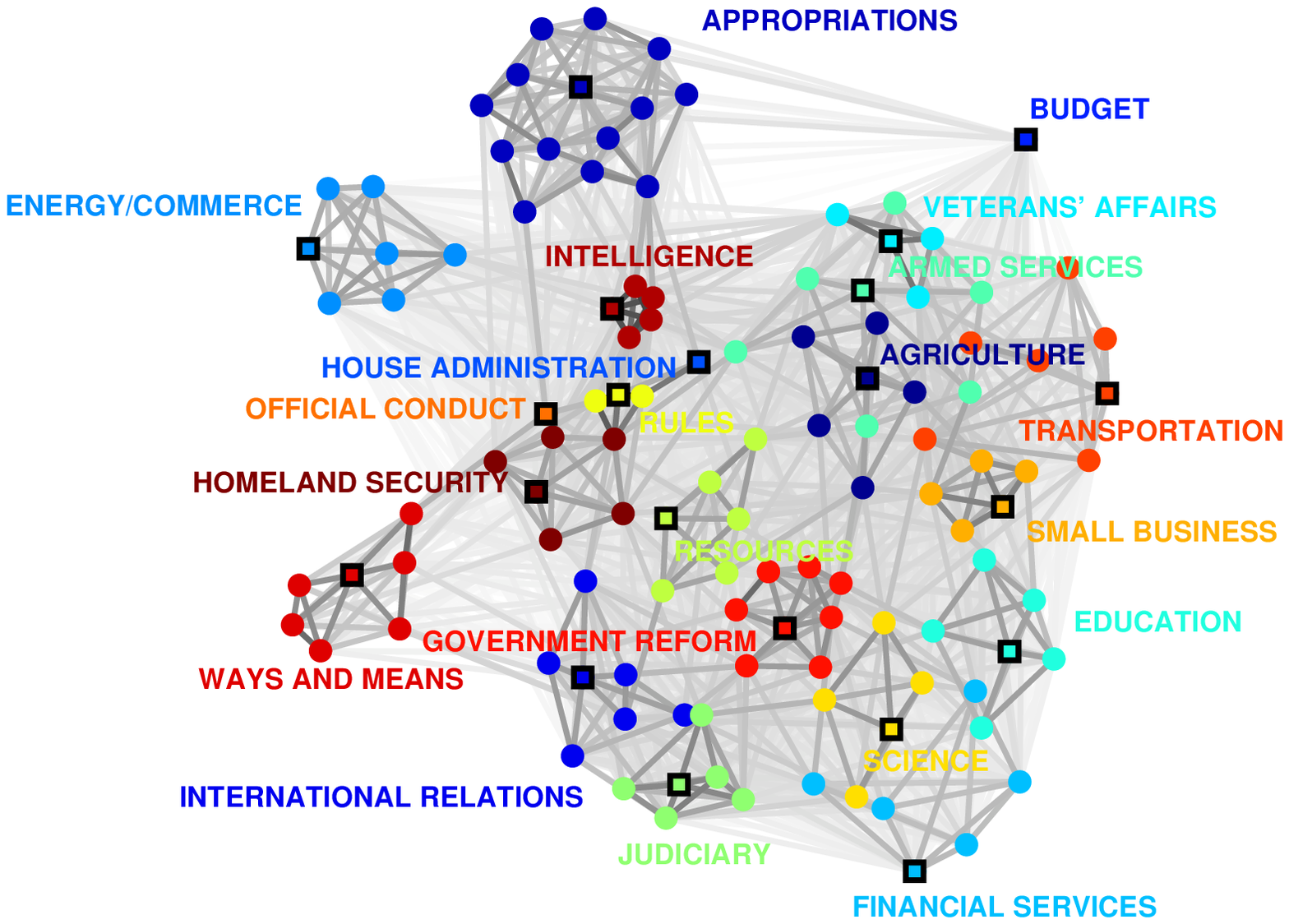}
\includegraphics[width = .5\textwidth]{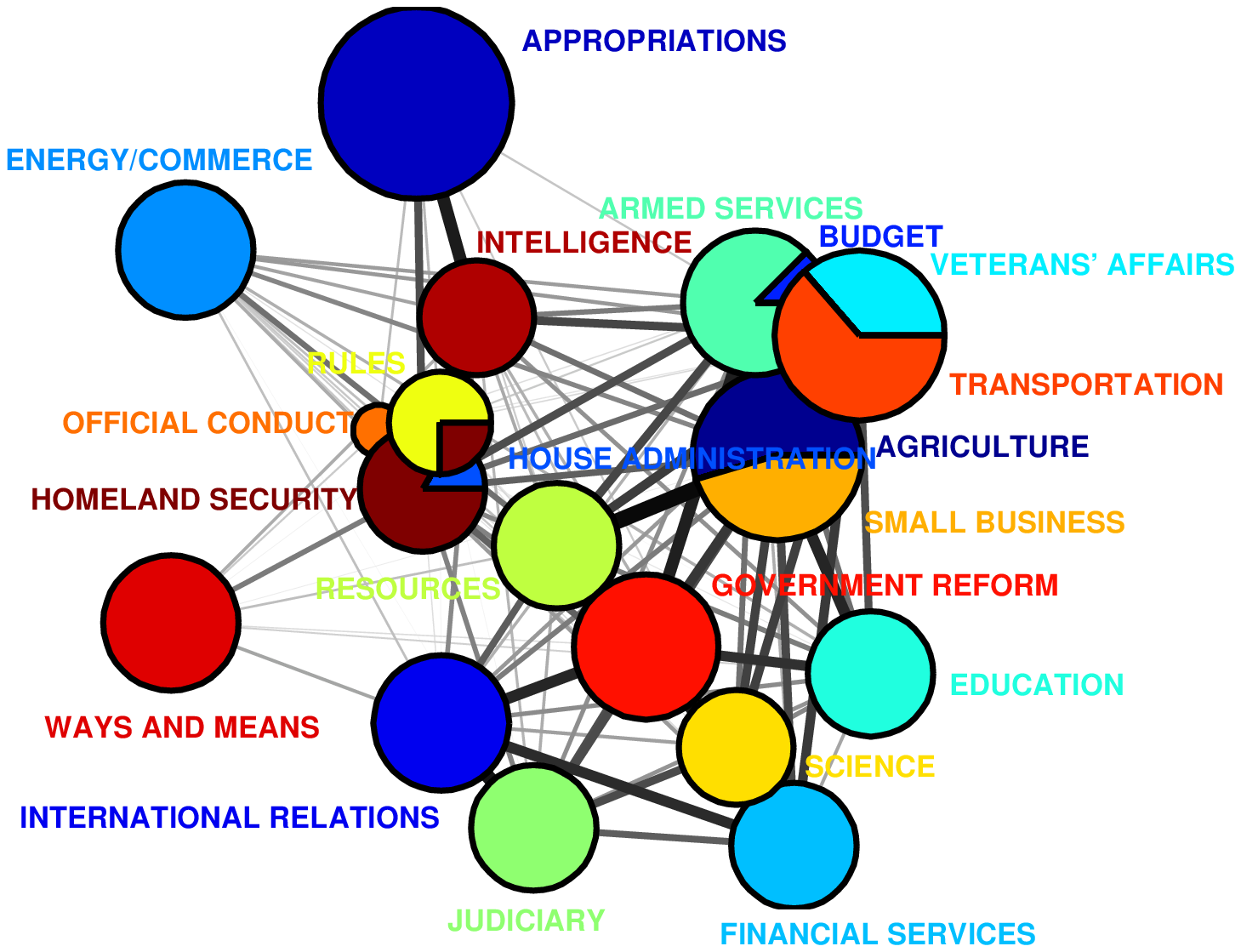}
}
\caption{(Left) The network of committees (squares) and subcommittees (circles) in the 108th U.S.~House of Representatives (2003-04), color-coded by the parent standing and select committees and visualized using the Kamada-Kawaii method \cite{kk}.  The darkness of each weighted edge between committees indicates how strongly they are connected.  Observe that subcommittees of the same parent committee are closely connected to each other.  (Right) Coarse-grained plot of the communities in this network.  Here one can see some close connections between different committees, such as Veterans Affairs/Transportation and Rules/Homeland Security.}
\label{intro}
\end{figure}

Sociologists recognized early that they needed powerful mathematical tools and large-scale data manipulation to address this challenging problem. An important step was taken in 2002, when Michelle
Girvan and Mark Newman brought graph-partitioning problems to the broader attention of the statistical physics and mathematics communities \cite{structpnas}.  Suddenly, community detection in networks became hip among physicists and applied mathematicians all over the world, and numerous new methods were developed to try to attack this problem.  The amount of research in this area has become massive over the past seven years (with new discussions or algorithms posted on the arXiv preprint
server almost every day), and the study of what has become known as \text{community structure} is now one of the most prominent areas of network science \cite{santolong,satu07,commreview}.

Although a rigorous notion of community structure remains elusive, one can clarify some issues through the notions of \textit{modules} and \textit{hierarchies}.  In general, a network's
community structure encompasses a potentially complicated set of
hierarchical and modular components \cite{santolong,structpnas,herbert}.  In this
context, the term module is typically used to refer to a single
cluster of nodes.  Given a network that has been partitioned into non-overlapping
modules in some fashion (although some methods also allow for
overlapping communities), one can continue dividing each module in an
iterative fashion until each node is in its own singleton community.  This hierarchical
partitioning process can then be represented by a tree, or \textit{dendrogram} (see Fig.~\ref{karateplot}).  Such processes can yield a hierarchy of nested modules (see Fig.~\ref{clausetfigure}), or a collection of modules at one mesoscopic level might be obtained in an algorithm independently from those at another level.  However obtained, the \textit{community
structure} of a network refers to the set of graph partitions obtained at
each ``reasonable'' step of such procedures.  Note that community
structure investigations rely implicitly on using connected network components.  (We will assume such connectedness in our discussion of community-detection algorithms below.) 
Community detection can be applied individually to separate components of networks that are not connected.

Many real-world networks possess a natural hierarchy.  For example, the committee assignment network of the U.~S. House of Representatives includes the House floor, groups of
committees, committees, groups of subcommittees within larger
committees, and individual subcommittees \cite{congshort,conglong}.
As shown in Fig.~\ref{intro}, different House committees are resolved
into distinct modules within this network.  At a different
hierarchical level, small groups of committees belong to larger but
less densely-connected modules.  To give an example closer to home, let's consider the departmental organization at a university and suppose that the network in Fig.~\ref{clausetfigure} represents collaborations among professors.  (It actually represents grassland species interactions \cite{clausetnature}.)  At one level of inspection, everybody in the mathematics department might show up in one community, such as the large one in the upper left.  Zooming in, however, reveals smaller communities that might represent the department's subfields.

Although network community structure is almost always fairly complicated, several forms of it have nonetheless been observed and shown to be insightful in applications.  The structures of communities and between communities are important for the demographic
identification of network components and the function of dynamical
processes that operate on networks (such as the spread of opinions and
diseases) \cite{santolong}.  A community in a social network
might indicate a circle of friends, a community in the World Wide Web might
indicate a group of pages on closely-related topics, and a community in a cellular
or genetic network might be related to a functional module. In
some cases, a network can contain several identical replicas of
small communities known as motifs \cite{milo}.  Consider a transcription network that controls gene
expression in bacteria or yeast.  The nodes represent genes or
operons, and the edges represent direct transcriptional regulation.  A
simple motif called a ``feed-forward loop'' has been shown both
theoretically and experimentally to perform signal-processing tasks
such as pulse generation. Naturally, the situation becomes much more
complicated in the case of people (doesn't it always?). However, monitoring
electronically-recorded behavioral data, such as mobile phone calls,
allows one to study underlying social structures
\cite{jp,marta08}.  Although these pair-wise interactions (phone calls)
are short in duration, they are able to uncover social groups
that are persistent over time \cite{vicsek}. One interesting empirical
finding, hypothesized by Granovetter \cite{weak}, is that links
within communities tend to be strong and links between
them tend to be weak \cite{jp}.  This structural configuration
has important consequences for information flow in social systems \cite{jp} and thus affects how the underlying network channels the circulation of social and cultural resources.  (See below for additional discussion.)

With methods and algorithms drawn from statistical physics, computer
science, discrete mathematics, nonlinear dynamics, sociology, and
other subjects, the investigation of network community structure (and more general forms of data clustering) has captured the attention of a diverse group of scientists
\cite{santolong,suneoutlook,satu07,newmanphystoday}.  This breadth of
interest has arisen partly because the development of community-detection
methods is an inherently interdisciplinary endeavor and
partly because interpreting the structure of a community and its
function often requires application-specific knowledge. In fact, one
aspect that makes the problem of detecting communities so challenging is
that the preferred formulation of communities is often domain-specific.  Moreover, after choosing a formulation, one still has to construct the desired communities by solving an optimization problem that is known to be NP-complete in at least one family of formulations \cite{np}. This has necessitated the
adaptation of classical computational-optimization heuristics and the development of new ones.

\section*{A Simple Example}\label{simple}

To set the stage for our survey of community-detection algorithms below,
consider the ubiquitous but illustrative example of the Zachary Karate Club, in which an
internal dispute led to the schism of a karate club into two smaller clubs \cite{karate}.  We show a visualization of the friendships between members of the original club in
Fig.~\ref{karateplot}. When the club split in two, its members chose preferentially to be in the one with most of their friends.  Sociologist Wayne Zachary, who was already studying the club's friendships when the schism occurred, realized that he might have been able to predict the split in advance.  This makes the Karate Club a useful benchmark for community-detection algorithms, as one expects any algorithmically-produced division of the network into communities to include groups that are similar to the actual memberships of the two smaller clubs.

In Fig.~\ref{karateplot}, we show the communities that we obtained using a spectral partitioning optimization of a \textit{quality function} known as \textit{modularity} \cite{newmodlong}.  (This method is described below.)
Keeping in mind the hierarchical organization that often occurs as part of network community structure, we visualize the identified divisions using a polar-coordinate dendrogram and enumerate the network's nodes around its exterior.  Each distinct radius of the dendrogram corresponds to a partition of the original graph into multiple groups.  That is, the community assignments at a selected level of the dendrogram are indicated by a radial cut in the right panel of Fig.~\ref{karateplot}; one keeps only connections (of nodes to groups) that occur outside this cut. 
The success of the community identification is apparent in the Karate Club example, as the two main branches in the dendrogram reflect the actual memberships of the new clubs.  

As shown in Fig.~\ref{karateplot}, this community-detection method subsequently splits each of the two main branches.  Hence, we see that the Zachary Karate Club network has a natural hierarchy of decompositions: a coarse pair of communities that correspond precisely to the observed membership split, and a finer partition into four communities.  In larger networks, for which algorithmic methods of investigation are especially important, the presence of multiple such partitions indicates mesoscopic network structures at different mesoscopic resolution levels.  At each level, one can easily compare the set of communities with identifying characteristics of the nodes (e.g., the post-split Karate Club memberships) by drawing a pie chart for each community, indicating the composition of node characteristics in that community, and showing the strength of inter-community connections as ties between the pies (as in Fig.~\ref{intro} for Congressional committees).


\section*{Identifying Communities}\label{methods}

Intuitively, a community is a cohesive group of nodes that are connected ``more densely" to each other than to the nodes in other communities. The differences between many community-detection methods ultimately come down to the precise definition of ``more densely" and the algorithmic heuristic followed to identify such sets. As different scientific fields have different needs, it is not surprising that a wide variety of community-detection methods have been developed to serve those needs \cite{santolong}. These differing needs have also resulted in the deployment of different real and computer-generated benchmark networks to test community-finding algorithms \cite{santolong,santo3}.  A 2005 review article \cite{commreview} compared the performance of several of the (then-)available methods in terms of both computation time and output.  A thorough, more recent discussion is available in Ref.~\cite{santolong}.

Rather than attempt a similar comparison using every available algorithm, our aim is instead to expose a larger readership to many of the most popular methods (as well as a few of our personal favorites), while contrasting their different perspectives and revealing a few important (and sometimes surprising) similarities.  Although we will attempt to highlight an extensive suite of techniques in our survey below, there are of course numerous other methods---including ones based on maximum likelihood \cite{clausetnature}, mathematical programming \cite{kempe}, block modeling \cite{hw76,white07}, link partitioning \cite{evans,sunelink}, inference methods \cite{hastings,leichtpnas}, latent space clustering \cite{handcock2007}, and more---that we unfortunately neglect here because of space considerations.   Many of them are discussed in other review articles \cite{santolong,satu07,commreview}.

\subsection*{Traditional Clustering Techniques}

The idea of organizing data by coarse-graining according to common
features is a very old one \cite{santolong,slater}.  The original
computational attempts to find clusters of similar objects are rooted in statistics and data mining.  Important methods include \textit{partitional clustering} techniques such as
$k$-means clustering, \textit{neural network clustering} techniques such as self-organizing maps, and \textit{multi-dimensional scaling} (MDS) techniques such as singular value
decomposition (SVD) and principal component analysis (PCA) \cite{siamcluster}.  For example, MDS algorithms of various levels of sophistication have proven to be amazingly successful at
finding clusters of similar data points in myriad applications, such
as voting patterns of legislators and Supreme Court justices
\cite{pr,sirovich,congshort,conglong}.  Such techniques start with a matrix that indicates similarities (e.g., a tabulation of how every legislator voted on every bill) and return a coordinate matrix that minimizes an appropriate loss function.  In the U.~S.~Congress, this allows one to see that the most important dimensions correspond to the liberal-conservative axis (``partisanship") and how well a given legislator plays with others (``bipartisanship").  During periods of heightened racial tension, such analyses have also revealed a third dimension corresponding to the division between North and South \cite{pr}.

Another prominent set of classical techniques to detect cohesive groups in graphs are hierarchical clustering algorithms such as the \textit{linkage clustering} methods used in phylogenetic biology \cite{cluster,siamcluster}.  One starts with the complete set of $N$ individual nodes in a weighted network, represented by an \textit{adjacency matrix} $A$ whose elements (links) $A_{ij}$ indicate how closely nodes $i$ and $j$ are related to each other.  For the purpose of our presentation, we will only consider undirected networks, which implies that $A$ is symmetric (a few algorithms can also handle directed networks \cite{santolong,guim07,leicht08}).  Linkage clustering is an example of an \textit{agglomerative} method, as it starts from individual nodes and ultimately connects the entire graph.
The nodes are conjoined sequentially into larger clusters, starting with the pair with maximal $A_{ij}$ (i.e., the most strongly connected pair).  At each step, one recomputes the similarities between the new cluster and each of the old clusters and again joins the two maximally-similar clusters, and one continues iteratively until all clusters with nonzero similarity are connected.  Different linkage clustering methods utilize different measures of the similarity between clusters.  For instance, in \textit{single linkage clustering}, the similarity of two clusters $X$ and $Y$ is defined as the greatest similarity between any pair of nodes $x \in X$ and $y \in Y$.  Joining nodes using single linkage clustering essentially mirrors Joseph Kruskal's algorithm for computing minimum spanning trees (MSTs) \cite{gower,eisner}.  With clustering, however, the order of cluster formation is important and can be represented as a dendrogram, whose depths indicate the steps at which two clusters are joined.  More sophisticated techniques that build on these ideas are discussed in Ref.~\cite{siamcluster}.

There are also a few classical \textit{divisive} techniques, in which one starts with the full graph and breaks it up to find communities \cite{santolong,siamcluster,satu07}.  (As with agglomerative techniques, one can visualize the results using dendrograms.)  The most prominent examples are spectral methods, which we discuss in detail below.
New data clustering methods, which are applicable both to networks and to more general data structures, continue to be developed very actively \cite{siamcluster,satu07}.  Scientists studying ``community detection" and those studying ``data clustering" are obviously looking at the same coin.  The two fields are advancing in parallel, and there are numerous deep connections between the two (including, we suspect, far more than are already known).

\subsection*{The Kernighan-Lin Algorithm} \label{lin}

An algorithm from computer science, which can be combined with other methods, was proposed by Brian Kernighan and Shen Lin (KL) in 1970 in their study of how to partition electric circuits into boards so that the nodes in different boards can be linked to each other using the fewest number of connections \cite{kl}.  To do this, they maximized a quality function $\tilde{Q}$ that relates the number of edges inside each group of nodes to the number between different groups.  Starting with an initial partition of a graph into two groups of predefined size, KL steps swap subsets containing equal numbers of vertices between the two groups.  To reduce the chance of getting stuck at a local maximum, the KL method permits swaps that decrease $\tilde{Q}$.  After a specified number of swaps, one keeps the partition with maximal $\tilde{Q}$ to use as the initial condition for a new set of KL steps.  When the number and sizes of communities are not specified, a natural generalization of the KL method is to move a single node at a time \cite{newmod,newmodlong,blondel,richardson}. Unsurprisingly, the partitions of networks into communities that are obtained using the KL algorithm depend strongly on one's initial partition and, therefore, it is best used as a supplement to high-quality partitions obtained using other methods \cite{newmod,newmodlong,santolong}.  In typical situations, both the KL swaps and the other method would seek to optimize the same $\tilde{Q}$.

\subsection*{Centrality-Based Community Detection}

Michelle Girvan and Mark Newman generated greater attention in mathematics and statistical physics for network community structure in Ref.~\cite{structpnas} when they devised a community-finding algorithm based on the sociological notion of \textit{betweenness centrality} \cite{antho,linton77,faust}.  An edge has high betweenness if it lies on a large number of paths between vertices.  (Note that betweenness can also be defined for nodes.)  If one starts at a node and wants to go to some other node in the network, it is clear that some edges will experience a lot more traffic than others.  The betweenness of an edge quantifies such traffic by considering strictly shortest paths (\textit{geodesic betweenness}) or densities of random walks (\textit{random walk betweenness}) \cite{walkbetween} between each pair of nodes and averaging over all possible pairs.  One can identify communities through a process of ranking each of the edges based on their betweenness, removing the edge with the largest value, and recalculating the betweenness for the remaining edges.  The recalculation step is important because the removal of an edge can cause a previously low-traffic edge to have much higher traffic.  An iterative implementation of these steps gives a divisive algorithm for detecting community structure, as it deconstructs the initial graph into progressively smaller connected chunks until one obtains a set of isolated nodes.  

Betweenness-based methods have been generalized to use network components other than edges, to bipartite networks \cite{conglong}, and to use other sociological notions of centrality \cite{santolong}.  However, although centrality-based community detection is intuitively appealing, it can be too slow for many large networks (unless they are very sparse) and tends to give relatively poor results for dense networks.

\subsection*{$k$-Clique Percolation and other Local Methods}

The method of \textit{$k$-clique percolation} \cite{vicsek} is based on the concept of a \textit{$k$-clique}, which is a complete subgraph of $k$ nodes that are connected with all $k(k-1)/2$ possible links. The method relies on the observation that communities seem to consist of several small cliques that share many of their nodes with other cliques in the same community. A \textit{$k$-clique community} is then defined as the union of all ``adjacent'' $k$-cliques, which by definition share $k-1$ nodes. One can also think about ``rolling'' a $k$-clique template from any $k$-clique in the graph to any adjacent $k$-clique by relocating one of its nodes and keeping the other $k-1$ nodes fixed \cite{palla05b}.  A community, defined through the percolation of such a template, then consists of the union of all subgraphs that can be fully explored by rolling a $k$-clique template.  As $k$ becomes larger, the notion of a community becomes more stringent.  Values of $k=3,\ldots,6$ then to be most appropriate because larger values become unwieldy.  The special case of $k = 2$ reduces to bond (link) percolation and $k = 1$ reduces to site (node) percolation.
 
The $k$-clique percolation algorithm is an example of a \textit{local} community-finding method.  One obtains a network's global community structure by considering the ensemble of communities obtained by looping over all of its $k$-cliques.  Some nodes might not belong to any community (because they are never part of any clique), and others can belong to several communities (if they are located at the interface between two or more communities).  The nested nature of communities is recovered by considering different values of $k$, although $k$-clique percolation can be too rigid because focusing on cliques typically causes one to overlook other dense modules that aren't quite as well-connected.  On the other hand, the advantage of $k$-clique percolation and other local methods is that they have to date provided one of the most successful ways to consider community overlap.

Allowing the detection of network communities that overlap is especially appealing in the social sciences, as people belong simultaneously to several communities (colleagues, family, hobbies, etc.)
\cite{white01,moody03}. Purely agglomerative or divisive techniques do not allow communities to overlap, so it is important to consider local methods as well.  Several such methods have now been developed \cite{localreview,santolong,wuhuberman,bagrow,clausetlocal,palla05b,evans,sunelink,borgatti90,luce49,seidman78}, including one that enables the consideration of overlapping communities at multiple resolution levels \cite{santo2}.  We believe further development of global clustering algorithms that take community overlap explicitly into account is essential to complement the insights from these local approaches.

\subsection*{Modularity Optimization}

One of the most popular quality functions is \textit{modularity}, which attempts to measure how well a given partition of a network compartmentalizes its communities \cite{structmix,structeval,markfast,newmod,newmodlong}.  The problem of optimizing modularity is equivalent to an instance of the famous MAX-CUT problem \cite{newmodlong}, so it is not surprising that it has been proven to be NP-complete \cite{np}.  There are now numerous community-finding algorithms that try to optimize modularity or similarly-constructed quality functions in various ways \cite{santolong,commreview,arenas08,blondel}.  In the original definition of modularity, an unweighted and undirected network that has been partitioned into communities has modularity \cite{structeval,markfast}
\begin{equation}
  	Q = \sum_i (e_{ii} - b_i^2)\,, \label{mod}
\end{equation}
where $e_{ij}$ denotes the fraction of ends of edges in group~$i$ for which the other end of the edge lies in group~$j$, and $b_i = \sum_j e_{ij}$ is the fraction of all ends of edges that lie in group~$i$.  Modularity is closely related to the Freeman segregation index \cite{freeman78}; a key difference is that $Q = 0$ when all nodes are assigned to the same community, which enforces the existence of a nontrivial partition with $Q > 0$.  Modularity explicitly takes degree heterogeneity into account, as it measures the difference between the total fraction of edges that fall within groups versus the fraction one would expect if edges were placed at random (respecting vertex degrees).\footnote{Interest in degree heterogeneity exploded in the late 1990s with the sudden wealth of empirical data and the seemingly ubiquitous manifestation of heavy-tailed degree distributions such as power laws \cite{ba2002,newmansirev}.}  Thus, high values of $Q$
indicate network partitions in which more of the edges fall within groups than expected by chance (under a specified null model, as discussed further below).  This, in turn, has been found to be a good indicator of functional network divisions in many cases~\cite{newmod,newmodlong}.

For weighted networks, one counts the sums of the weights of edges instead of the number of edges, so heavily-weighted edges contribute more than lightly-weighted ones. Both $e_{ij}$ and $b_i$ are thus straightforwardly generalized, and then the modularity is again calculated from
Eq.~(\ref{mod}).  The meaning of modularity remains essentially the
same: It measures when a particular division of the network has more edge
weight within groups than one would expect by chance.

Quality functions such as modularity provide precise statistical measures of how to count the total strength of connections within communities versus those between communities
\cite{structmix,newmod}.  Modularity is a scaled \textit{assortativity} measure based on whether high-strength edges are more or less likely to be adjacent to other high-strength edges \cite{structmix,structeval,markfast}.  Because communities are supposed to have high edge density relative to other parts of the graph, a high-modularity partition tends to have high edge-strength assortativity by construction.  More generally, assortativity notions can be used to
partition a graph into groups according to any characteristic by examining whether
nodes are more likely (in \textit{assortative} graphs) or less likely (in \textit{disassortative graphs}) to be connected to nodes of the same type \cite{newmansirev}.  

Interestingly, maximizing modularity is closely related to the energy models of pairwise attraction, such as the Fruchterman-Reingold method, that are commonly used for graph visualization \cite{noack08}.  
While this isn't necessarily surprising given the clusters that one can typically observe with good graph visualization tools, this recent insight does suggest that such tools may also help lead to better community-detection methods.  Conversely, the analysis and construction of algorithms to find network communities might help lead to better graph-visualization techniques.

It is typically impossible computationally to sample a desired quality function by exhaustively enumerating the non-polynomial number of possible partitions of a network into communities \cite{np}. A number of different methods have thus been proposed to balance the typical quality of their identified optima with the computational costs.  Some methods, such as the greedy algorithms in Refs.~\cite{markfast,clausetfast}, are fast heuristics intended to be applied to networks with millions of nodes or more.  Other methods---such as spectral partitioning \cite{newmod,newmodlong} (discussed below), refined greedy algorithms \cite{schuetz}, simulated annealing \cite{amaral}, extremal optimization \cite{duch}, and others \cite{noack08b}---provide more sophisticated but slower means to identify high-modularity partitions.\footnote{As we have discussed, one can also supplement any of these methods with KL swapping steps \cite{kl,newmod,newmodlong,blondel,richardson,noack08b}.} We further discuss the spectral partitioning method below, in part because of its interesting reformulation of the modularity scalar as a matrix, but we note that other algorithmic choices may be superior in many situations.  We believe that there is considerable value in having multiple computational heuristics available, as this provides greater flexibility to compare and contrast the identified communities.

Importantly, many modularity-maximization techniques are easy to generalize for use with other related quality functions because it is far from clear that modularity is the best function to optimize.  For example, modularity has a known \textit{resolution limit} (see below)
that might cause one to miss important communities \cite{resolution}.  A few alternatives to modularity have been considered \cite{santolong,signed,santo2,wiggy08,arenas08}, and it is ultimately desirable to optimize a quality function that includes not only network structure but also other information (such as node characteristics or relevant time-dependence) that would allow one to incorporate functionality directly \cite{cosma}.  Such consideration of additional information is one of the most important open issues in community detection \cite{newmanphystoday,facebook}.

\subsection*{Spectral Partitioning} \label{spectral}

The method of spectral partitioning arose most prominently in the development of algorithms for parallel computation \cite{fiedler,pothen}.  In traditional spectral partitioning, network properties are related to the spectrum of the graph's Laplacian matrix $L$, which has components 
$L_{ij} = k_i\delta(i,j) - A_{ij}$, where $k_i$ is the degree of node $i$ (or, in a weighted network, its strength), and $\delta(i,j)$ is the Kronecker delta (i.e., $\delta(i,j)=1$ if $i = j$, and $0$ otherwise).  

The simplest such method starts by splitting a network into two components.  One then applies two-group partitioning recursively to the smaller networks one obtains as long as it
is desirable to do so.  (One can also partition networks into more than two groups during each step \cite{capocci,donetti,newmodlong,richardson}.)  For a single partitioning step, one defines an index
vector $s$ whose components take the value $+1$ if they belong to
group $1$ and $-1$ if they belong to group $2$.  The matrix of edges between each pair of nodes can then be expressed as $R = \frac{1}{4}s^TLs$. 
The ``best'' partition of the network seemingly results from choosing $s$
to minimize $R$ (called the ``cut size") and hence the total strength of edges between 
the two groups.  (Recall the max-flow min-cut theorem, which states that the minimum cut between any
two vertices of a graph---that is, the minimum set of edges whose deletion places the two vertices in disconnected components of the graph---carries the maximum flow between the two vertices\cite{ford56,elias56}.)  Unfortunately, this minimization is easily accomplished by choosing the trivial (and useless) partition of a single group containing every node. The most common solution to this situation is to fix the sizes of the two groups in advance and incorporate this information in the partitioning procedure (as described in, e.g., \cite{newmodlong}).  This solution is perfectly reasonable for some applications, such as load balancing in parallel computing.  However, this approach is neither appropriate nor realistic for community detection in most other contexts because one typically does not know the number or sizes of communities in advance, so choosing arbitrary sizes at the outset precludes attacking the main problem of interest.  

Fortunately, one can use the idea of modularity to obtain spectral partitioning algorithms that are appropriate for a broader class of problems \cite{newmod} (see also the earlier publication \cite{donetti} and a spiritually similar approach based on peer influences in the sociology literature \cite{moody01}).  By reformulating the scalar quantity of modularity in terms of a \textit{modularity matrix} $B$, with components 
  \begin{equation}
	B_{ij} = A_{ij} - P_{ij}\,, \label{moduse}
\end{equation}
spectral partitioning can be directly applied \cite{newmod} as a means of heuristically optimizing the modularity
\begin{equation}
  Q = \frac{1}{2W} \sum_{i,j} B_{ij} \delta(C_i,C_j)\,,
\end{equation}
where $\delta(C_i,C_j)$ indicates that the $B_{ij}$ components are only summed over cases in which nodes $i$ and $j$ are classified in the same community.  The factor $W = \frac{1}{2}\sum_{ij}A_{ij}$ is the total edge strength in the network (equal to the total number of edges for unweighted networks), where $k_{i}$ again denotes the strength of node $i$.  In (\ref{moduse}), $P_{ij}$ denotes the components of a  \textit{null model} matrix, which specifies the relative value of intra-community edges in assessing when communities are closely connected \cite{newmodlong,bansal08}. In general, one is free to specify any reasonable null model. The most popular choice, proposed by Newman and Girvan \cite{structmix,structeval,newmod,newmodlong}, is
\begin{equation}
	P_{ij} = \frac{{k_i k_j }}{{2W}}\,. \label{ngnull}
\end{equation}
This recovers the definition of modularity in Eq.~(\ref{mod}), specified in terms of edge-weight deviations from a network chosen randomly from the set of all graphs with the same expected strength distribution as the actual network.  This null model is closely related to the configuration model \cite{molloy1995}, which (as with Erd\"os-Renyi random graphs) yields networks that aren't expected to have a natural hierarchy \cite{molloy1995,newmansirev,structeval}  The difference is that (\ref{ngnull}) is conditioned on the expected degree (or strength) sequence, whereas the configuration model is conditioned on the actual observed sequence.

In spectral partitioning, one can use as many eigenvectors of $B$ as there are positive eigenvalues,
but it is effective (and simplest) to recursively subdivide a network using only the ``leading eigenvector'' $v$, which is paired with the largest positive eigenvalue of $B$. One can then separate the network into two communities according to the signs $s_i=\mathrm{sgn}(v_i)$.  The magnitude of $v_i$ indicates the strength to which the $i$th node belongs to its assigned community \cite{newmodlong}.  For $v_i = 0$, one can assign node $i$ to a community based on which choice produces the higher modularity, changing $s_i = 0$ to $+1$ or $-1$ as appropriate.  The modularity of the resulting two-group partition of the network is $Q = \frac{1}{4W}s^TBs$.  After this bipartition, one then repeats this procedure for each graph component, keeping track of the fact that they are actually part of a larger network.  One continues recursively until the modularity can no longer be increased with additional subdivisions
\cite{newmod,newmodlong}.  The final network partition gives the community structure at a specific resolution level (e.g., committees in the U.~S. House of Representatives committee assignment
network).  This method can be generalized by considering different quality functions \cite{santolong,richardson}, allowing steps that decrease global quality in order to further subdivide the communities \cite{yan,richardson}, using more eigenvectors \cite{vectorwang,richardson}, or including a resolution parameter \cite{spinglass,arenas08} that allows one to examine the network's community structure at different mesoscopic scales.

\subsection*{The Potts Method} \label{pottsmethod}

Particles that possess a magnetic moment are often called \textit{spins} \cite{schwabl,potts}.  Such spins interact with other spins either \textit{ferromagnetically} (they seek to align) or \textit{antiferromagnetically} (they seek to have different orientations). A \textit{spin glass} is a system that encompasses both disorder and competing ferromagnetic and antiferromagnetic interactions.  This leads to a very large number of metastable spin configurations separated by energy barriers with long, glass-like characteristic relaxation times \cite{potts, fischer1993}. An important recent insight, inspired by earlier work on data clustering based on the physical properties of an inhomogeneous ferromagnetic model \cite{blatt}, is that optimizing modularity is mathematically equivalent to minimizing the energy (i.e., finding the ground state of the Hamiltonian) of an infinite range $q$-state Potts model \cite{rb04,spinglass}.  

In a $q$-state Potts spin glass, each spin can have one of $q$ states. The interaction energy between spins $i$ and $j$ is given by $-J_{ij}$ if the spins are in the same state and zero if they are not \cite{potts,spinglass}.  The Hamiltonian of the system is given by the sum over all of the pairwise interaction energies:
\begin{equation}
	H(\{\sigma\}) = - \sum_{ij} J_{ij} \delta(\sigma_i, \sigma_j)\,, \label{pottsequ}
\end{equation}
where $\sigma_l$ indicates the state of spin $l$ and $\{\sigma\}$ denotes the configuration of spins (i.e. the state of each of the system's $N$ spins).  There are a total of $q^N$ such configurations.

We map the problem of minimizing (\ref{pottsequ}) to network community detection by assigning a spin to each node and letting $q = N$. In this language, one adds the interaction energy $-J_{ij}$ if and only if nodes $i$ and $j$ are placed in the same community. Two nodes that are connected by an edge interact ferromagnetically ($J_{ij} > 0$) when the weight of the edge is greater than expected (given a specific null model) and interact antiferromagnetically ($J_{ij} < 0$) when it is less than expected. If $J_{ij} = 0$, spins $i$ and $j$ do not interact with each other.  Hence, two nodes want to be in the same community if they interact ferromagnetically and in different ones if they interact antiferromagnetically. One cannot in general find any arrangement of spins (i.e., any partition of nodes into communities) that simultaneously minimizes all of the pairwise interaction energies. Regardless of this inability to satisfy all of the bonds simultaneously, a phenomenon termed ``frustration'' \cite{fischer1993}, one can still try to minimize $H$ globally to find the ground state of the system.  The choice of interaction matrix elements given by
\begin{equation}
	J_{ij} = \frac{A_{ij} - P_{ij}}{W} \label{pottsmod}
\end{equation}	
implies that $H = -Q$ and recovers modularity maximization \cite{spinglass}.  (Division by $W$ is a normalization and does not affect the optimization algorithms.)  Alternative interaction models can also be used to partition networks (see, e.g., \cite{wiggy08}).

\subsection*{Resolution Parameters} \label{reso}

In 2007, Santo Fortunato and Marc Barth\'{e}lemy demonstrated using both real and computer-generated networks that modularity suffers from a resolution limit in its original formulation \cite{resolution}, as it misses communities that are smaller than a certain threshold size that depends on the size of the network and the extent of interconnectedness of its communities.  Communities smaller than the threshold tend to be merged into larger communities, thereby missing important structures. We have seen this in our own work on the U.~S. House committee assignment network, as detecting communities by maximizing modularity typically groups multiple standing committees (with their subcommittees) into a single community \cite{conglong}.

One way to address this resolution limit is to incorporate an explicit \textit{resolution parameter} directly into equations like (\ref{pottsmod}) to obtain \cite{spinglass}
\begin{equation}
	J_{ij} = \frac{A_{ij} - \lambda P_{ij}}{W}\,. \label{withreso}
\end{equation}	
One can alternatively incorporate a resolution parameter into $A_{ij}$ or elsewhere in the definition of a quality function (see, e.g., \cite{arenas08}).  This allows one to zoom in and out in order to find communities of different sizes and thereby explore both the modular and the hierarchical structures of a graph.  Fixing $\lambda$ in \eqref{withreso} corresponds to setting the scale at which one is examining the network: Larger values of $\lambda$ yield smaller communities (and vice versa). Resolution parameters have now been incorporated (both explicitly and implicitly) into several methods that use modularity \cite{spinglass,blondel}, other quality functions \cite{santo2,arenas08}, and other perspectives \cite{bagrow,vicsek}.  

Although introducing a resolution parameter into equations like (\ref{withreso}) seems \textit{ad hoc} at first, it can yield very interesting insights.  For example, $J_{ij} = (A_{ij} - \lambda)/W$ gives a uniform null model in which a given fixed average edge weight occurs between each node.  This can be useful for correlation and similarity networks, such as those produced from matrices of yeah and nay votes. Nodes $i$ and $j$ want to be in the same community if and only if they voted the same way at least some threshold fraction of times that is specified by the value of $\lambda$.  

Even more exciting, one can relate resolution parameters to the time scales of dynamical processes unfolding on a network \cite{barahona1,barahona2,rosvall,latapy}.  Just as we can learn about the behavior of a dynamical system by studying the structural properties of the network on which it is occurring, we can also learn about the network's structural properties by studying the behavior of a given dynamical process. This suggests the intuitive result that the choice of quality function should also be guided by the nature of the dynamical process of interest.  In addition to revealing that resolution parameters arise naturally, this perspective shows that the Potts method arises as a special case of placing a continuous time random walk with Poisson-distributed steps on a network \cite{barahona2}.  Freezing the dynamics at a particular point in time yields the modularity-maximizing partition.  Freezing at earlier times yield smaller communities (because the random walker hasn't explored as much of the graph), and waiting until later times results in larger communities.  The $t \rightarrow \infty$ limit reproduces the partitioning from Miroslav Fiedler's original spectral method \cite{fiedler}.

\section*{Applications} \label{app}

Armed with the above ideas and algorithms, we turn to selected demonstrations of their efficacy.  The increasing rapidity of developments in network community detection has resulted in part from the ever-increasing abundance of data sets (and the ability to extract them, with user cleverness).  This newfound wealth---including large, time-dependent data sets---has, in turn, arisen from the massive amount of information that is now routinely collected on websites and by communication companies, governmental agencies, and others.  Electronic databases now provide detailed records of human communication patterns, offering novel avenues to map and explore the structure of social, communication, and collaboration networks.  Biologists also have extensive data on numerous systems that can be cast into network form and which beg for additional quantitative analyses.

Because of space limitations, we restrict our discussion to five example applications in which
community detection has played a prominent role: scientific coauthorship, mobile phone communication, online social networking sites, biological systems, and legislatures.  We make no attempt to be exhaustive for any of these examples; we merely survey research (both by others and by ourselves) that we particularly like.


\subsection*{Scientific Collaboration Networks} \label{author}

\begin{figure}
\centerline{
\includegraphics[width = 1.2\textwidth]{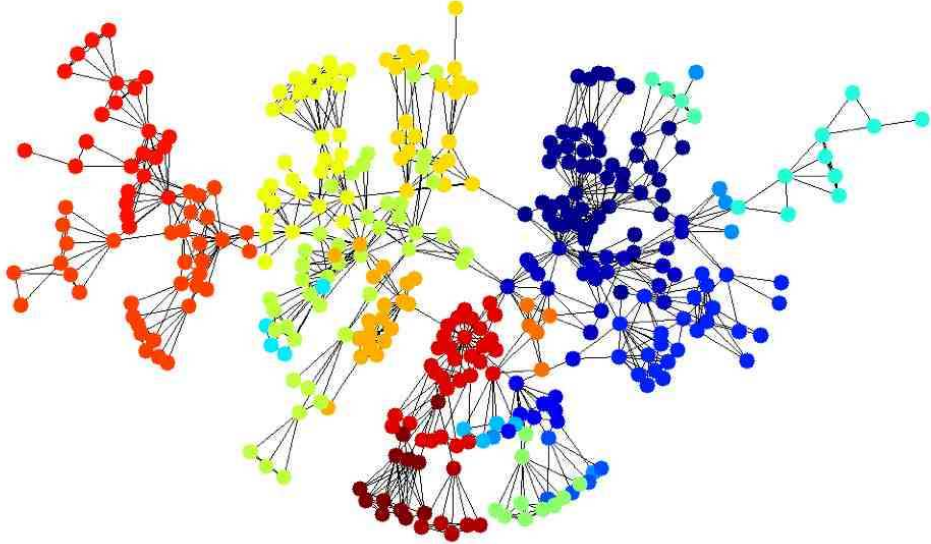}
\vspace*{-1in}}
\caption{The largest connected component (379 nodes) of the network of network scientists (1589 total nodes), determined by coauthorship of papers listed in two well-known review articles \cite{newmansirev,bocca06} and a small number of papers added manually \cite{newmodlong}.  Each of the nodes in the network, which we depict using a Kamada-Kawaii visualization \cite{kk}, is colored according to its community assignment using the leading-eigenvector spectral method \cite{newmodlong}. 
}
\label{meta}
\end{figure}

We know from the obsessive computation of Erd\"os numbers that scientists can be quite narcissistic. (If you want any further evidence, just take a look at the selection of topics and citations in this section.)
In this spirit, we use scientific coauthorship networks as our first example.  

A bipartite (two-mode) coauthorship network---with scientists linked to papers that they authored or coauthored---can be defined by letting $\delta_i^p=1$ if scientist $i$ was a coauthor on paper $p$ and zero otherwise. Such a network was collected and examined from different databases of research papers in Refs.~\cite{pnascollab,newmansc1,newmansc2}.  To represent the collaboration strength between scientists $i$ and $j$, one can then define 
\begin{equation}
	A_{ij} = \sum_p \frac{\delta_i^p\delta_j^p}{n_p-1} \label{uni}
\end{equation}
as the components of a weighted unipartite (one-mode) network, where $n_p$ is the number of authors of paper $p$ and the sum runs over multiple-author papers only.  Applying betweenness-based community detection to a network derived from Santa Fe Institute working papers using (\ref{uni}) yields communities that correspond to different disciplines \cite{structpnas}.  The statistical physics community can then be further subdivided into three smaller modules that are each centered around the research interests of one dominant member.  Similar results have been found using various community-finding algorithms and numerous coauthorship networks, such as the network of network scientists \cite{newmodlong} (see Fig.~\ref{meta}), which has become one of the standard benchmark examples in community-detection papers.


\subsection*{Mobile Phone Networks} \label{phone}

Several recent papers have attempted to uncover the large-scale characteristics of communication and social networks using mobile phone data sets \cite{jp,vicsek07,marta08}.  Like many of the coauthorship data sets studied recently \cite{vicsek07}, mobile phone networks are longitudinal (time-dependent).  However, in contrast to the ties in the coauthorship network above,
links in phone networks arise from instant communication events and capture relationships as they happen. This means that at any given instant, the network consists of the collection of ties connecting the people who are currently having a phone conversation.  To probe longer-term social structures, one needs to aggregate the data over a time window.

\begin{figure}
\centerline{
\includegraphics[width = 1\textwidth]{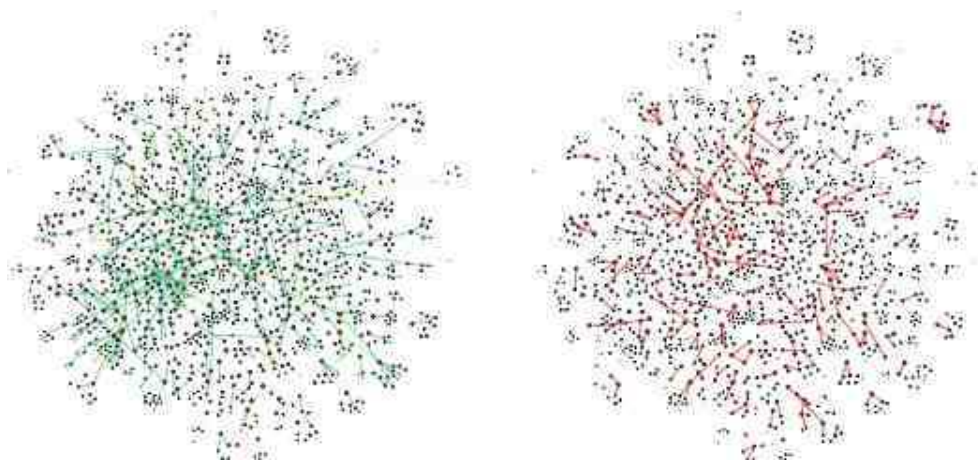}
}
\caption{(Left) A sample of a mobile phone network studied in Refs.~\cite{jp,jp2}. After the strong ties have been removed, the network still retains its global connectivity. (Right) Removal of weak ties leads, through a phase transition, to a disintegration of the network. Figure adapted from Ref.~\cite{jp2}.
}
\label{phonefig}
\end{figure}

In 2007, one research group used a society-wide communication network containing the mobile-phone interaction patterns of millions of individuals in an anonymous European country to explore the relationship of microscopic, mesoscopic, and macroscopic network structure to the strength of ties between individuals on a societal level \cite{jp}. They obtained very interesting insights into Mark Granovetter's famous \textit{weak tie hypothesis}, which states that the relative overlap of the friendship circle of two individuals increases with the strength of the tie connecting them \cite{weak}. At the mesoscopic level, this leads to a structure in which individuals within communities tend to be linked via strong ties, whereas communities tend to be connected to other communities via weak ties. Because of this coupling between link strength and function, the weak ties are responsible for the structural integrity of the communication network:  It is robust to the removal of the strong ties but breaks up if the weak ties are removed (see Fig.~\ref{phonefig}). In fact, one can even show that the removal of weak ties leads to a (phase) transition from a regime in which the network remains globally connected to one in which the network essentially consists of insular communities.  However, there is no phase transition if the strong ties are removed, so the network remains globally connected. The location of the transition also suggests a natural quantitative demarcation between weak and strong ties. This mesoscopic organization of social networks has important consequences for the flow of information. If one assumes that every tie (regardless of strength) is equally efficient in transferring information, one recovers the classical result of Granovetter that weak ties are mostly responsible for information diffusion \cite{weak}.  However, if one assumes that the rate of information transfer is proportional to the strength of the tie, then neither weak nor strong ties are as effective as intermediate ties for information diffusion \cite{jp}.

To help develop methods that can be applied to time-dependent networks, another research group has recently applied $k$-clique percolation to a large mobile phone data set to investigate community formation, evolution, and destruction \cite{vicsek07}.  They found that large communities persist longer if they are capable of dynamically altering their membership (suggesting that an ability to change the group composition results in better adaptability), whereas small groups last longer if they stay virtually unchanged.  We've all seen examples of such dynamics in real life:  A mathematics department at a university will last a long time and will still be identified as fundamentally the same community even though its membership can change quite drastically over time.  On the other hand, an individual research group might rely on only one or two of its members for its existence.

\subsection*{Online Social Networks} \label{face}

\begin{figure}
\centerline{
\includegraphics[width = .4\textwidth]{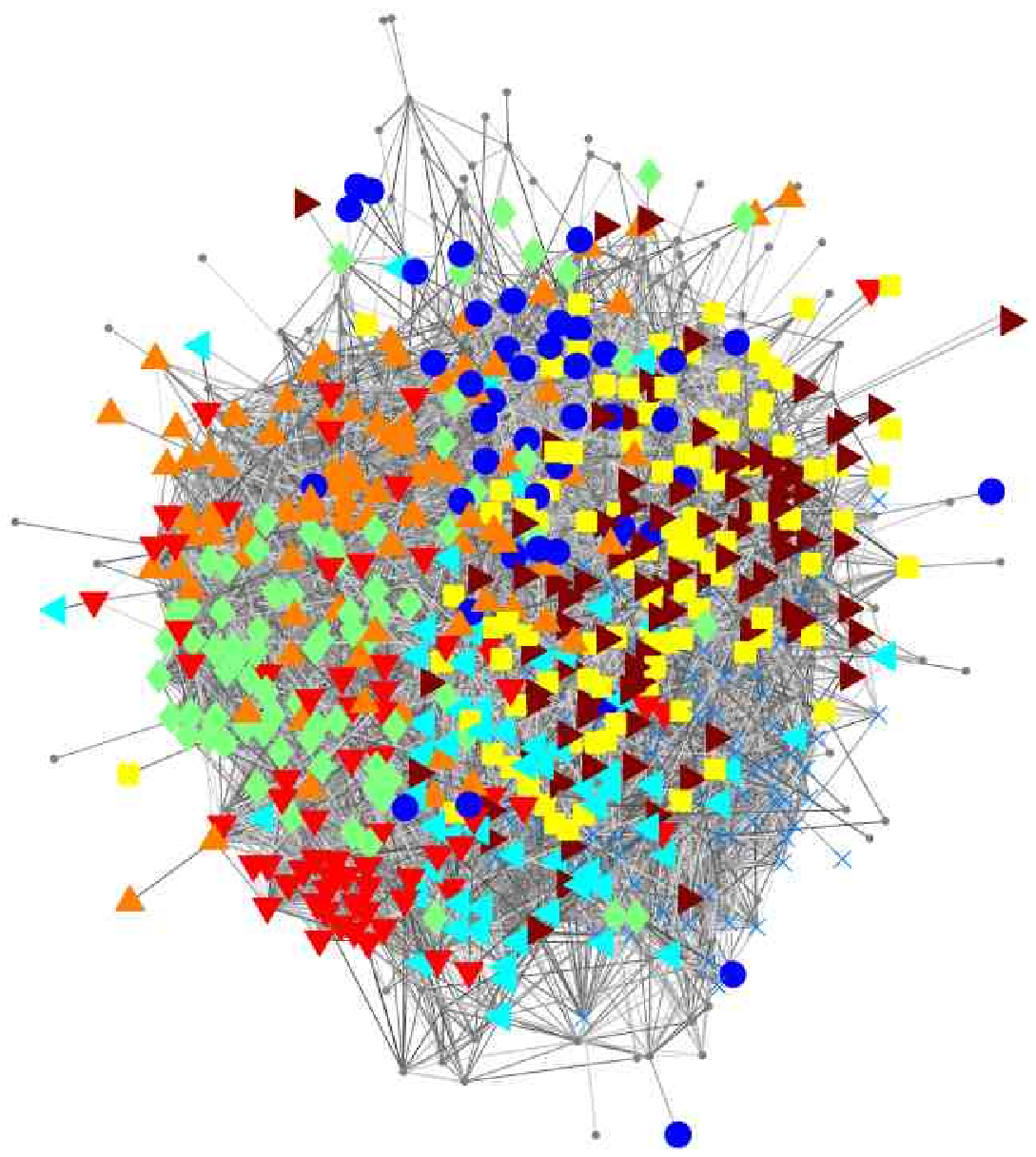}
\includegraphics[width = .6\textwidth]{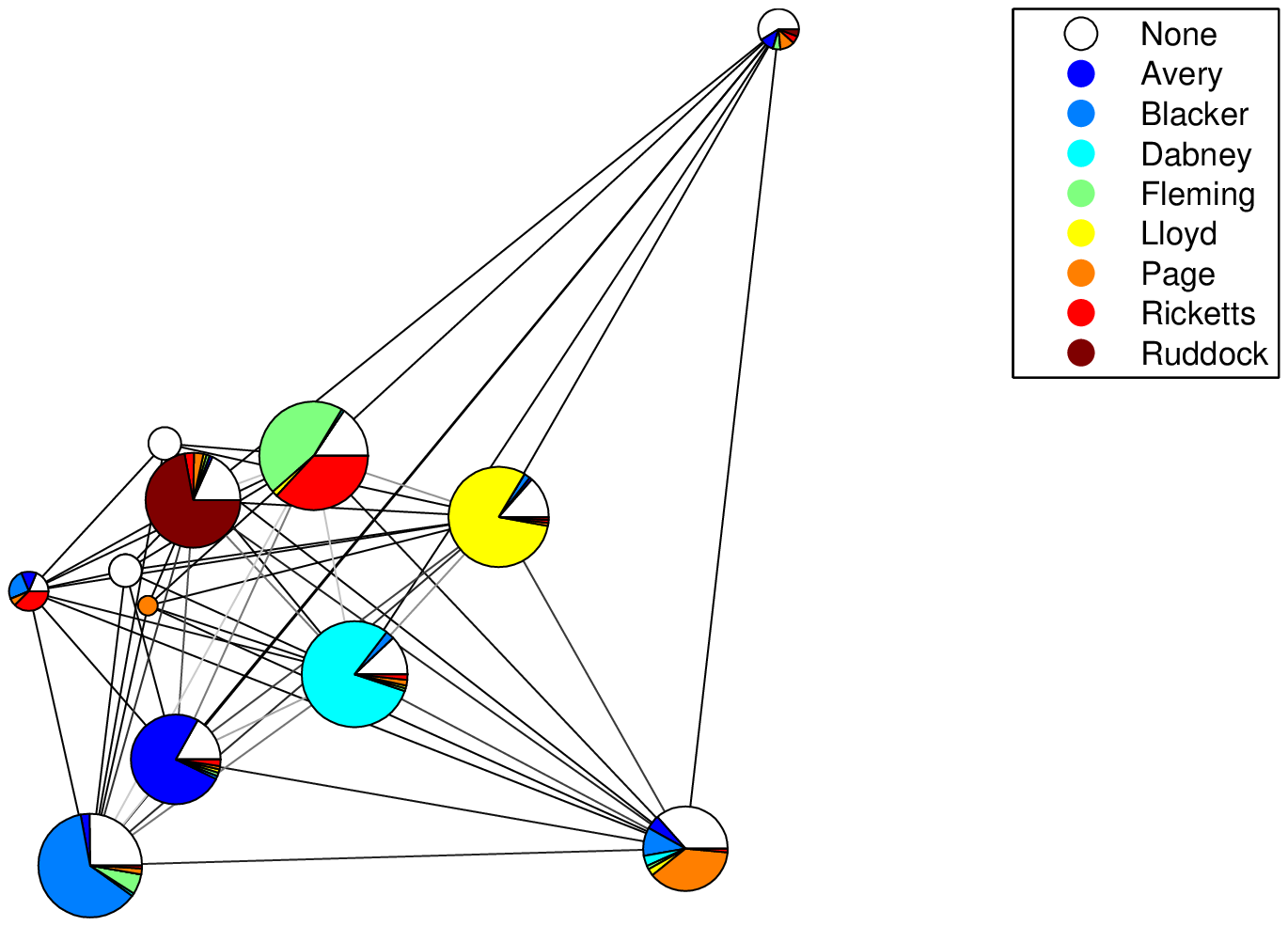}
}
\caption{(Left) Caltech friendship network based on Facebook data from September 2005 using the Fruchterman-Reingold \cite{fr91} visualization method.  The colors (and shapes) correspond to self-identified House (dormitory) affiliation.  (Right) Communities in the Caltech network.  We obtained this community structure, which accurately matches the Caltech House structure, using a slight modification of Newman's leading-eigenvector algorithm \cite{newmodlong} and subsequent KL node-swapping steps \cite{richardson,yan,facebook}. It also gives some indication of the most probable House affiliations of the people in white, who did not identify their House \cite{facebook}.   
}
\label{caltech}
\end{figure}

Social networking sites (SNSs) have become a pervasive part of everyday life. They allow users to construct a public or semi-public online profile within a bounded system, articulate a list of other users (called ``friends'') with whom they share a connection, and view and traverse their network of connections \cite{boydell,facebook}.  Since their introduction, SNSs such as Facebook, LinkedIn, MySpace, and hundreds of others have collectively attracted over one billion users \cite{wiki}.  People have integrated SNSs into their daily lives, using them to communicate with friends, send e-mails, solicit opinions or votes, organize events, spread ideas, find jobs, and more.

The emergence of SNSs has also revolutionized the availability of quantitative social and demographic data, and in turn transformed the study of social networks \cite{boydell}.  This has obviously been a big deal for social scientists (see Ref.~\cite{lewis} for an excellent recent example), but numerous mathematicians, computer scientists, physicists, and more have also had a lot of fun with this new wealth of data.  This has included investigations of attachment mechanisms to determine how SNS network
structure might develop \cite{kumar}, the formation of friends, allies, and nemeses  \cite{szabo1,szabo2}, and much more \cite{boydell,facebook}.

In a recent paper \cite{facebook}, we used anonymous Facebook data from September 2005 to compare the friendship networks of several American universities.  This data yields networks in which each node represents a person and each edge represents a self-identified, reciprocal online friendship.  The institutions we considered ranged from small technical universities such as Caltech (1089 nodes) to large state universities such as the University of Oklahoma (about 24000 nodes).  Our data also includes limited demographic information for each user: gender, high school, class year, major, and dormitory residence.  Using permutation tests, we developed tools that allow one to quantitatively compare different network partitions, which can be obtained from any desired method---including, in particular, community-detection algorithms and user demographics.  This enables one to investigate the demographic organization of different universities and compare the results of different community-detection algorithms.  We found, for example, that communities at Princeton, Georgetown, and the University of North Carolina at Chapel Hill are organized predominantly by class year, whereas those at Caltech are based almost exclusively on House (dormitory) affiliation.  As we illustrate in Fig.~\ref{caltech}, community structure can also be used to make simple yet intelligent guesses about withheld user characteristics.  Naturally, this opens up significant privacy issues when data is not fully anonymous.


\subsection*{Biological Networks} \label{bio}

One of the paramount goals of studying biological networks is to determine the principles governing their evolution.  It is hoped that finding important network structures might give some insights into the mechanisms (and, ideally, the natural design principles) that control the flow of biological information.  It is thus unsurprising that clustering methods form a core part of bioinformatics \cite{bower,kepes}, and there is simply no way to do justice to this vast literature here.  Accordingly, we only present a couple of our favorite examples.
 
In 2002, Ron Milo et al. investigated a plethora of directed networks to develop the idea of miniature communities known as \textit{motifs} \cite{milo}, which are used to describe patterns of nodes and edges that occur significantly more prevalently than expected in ensembles of random networks.  Motifs can be interpreted as basic building blocks of complex networks, perhaps representing small functional modules that arose via evolutionary mechanisms.  The simplest motifs---triangles (3-cliques), in which three nodes are all mutually interconnected---have long been studied in social networks \cite{faust,freemanbook}.  The amazing discovery of Milo et al. is that different types of motifs are, in fact, prevalent universally in many different types of networks.  Among the omnipresent motifs they observed are three-chains in ecological networks (in which a prey node is connected by a directed edge to a predator node, which is in turn connected by a directed edge to another predator); diamonds in ecological networks, neural networks, and logic chips; feed-forward loops in gene regulation networks, neural networks, and logic chips; fully-connected triangles in the World Wide Web; and more.

Numerous scientists have built on this foundation of motifs, and several of these investigations have provided fascinating connections between motifs and larger mesoscopic structures.  For example, one team investigated three-node and four-node motifs in an integrated \textit{Saccharomyces cerevisiae} network, which they constructed using information from protein interactions, genetic interactions, transcriptional regulation, sequence homology, and expression correlation \cite{gold05}.  Their primary finding was that most motifs form larger ``network themes" of recurring interconnection patterns that incorporate multiple motif occurrences.  They were also able to tie some of these mesoscopic themes to specific biological phenomena, such as a pair of protein complexes with many genetic interactions between them.  The notion of motifs has also recently been used to develop generalizations of graph modularity \cite{motifmod}.

One can imagine constructing a course-grained network consisting of interconnected network themes.  For example, in a seminal 2005 paper, Roger Guimer\`a and Lu\'{i}s Amaral used communities to try to
 construct a ``functional cartography" of biological networks in order to employ nodes and modules with known roles to try to obtain interesting insights on nodes and modules with unknown functionality \cite{amaral}.  To understand their perspective, consider the prototypical maps of countries in which important cities are marked by special symbols, other cities are marked with dots, and smaller towns are not shown.  In the network context, there is a one-to-one correspondence between markers and communities, and the symbols are determined according to structural and/or functional role.  The connections between communities are analogous to major highways.  In fact, our coloring of community pies according to the demographic composition of their nodes (see Figs.~\ref{intro} and \ref{caltech}) was originally inspired by Ref.~\cite{amaral}.

To illustrate their idea, Guimer\`a and Amaral considered the metabolic networks of twelve different organisms.  They started by detecting communities by maximizing modularity using simulated
annealing and then (following a suggestion from social scientist Brian Uzzi) calculated appropriate properties of each node to determine their role in their assigned community.  After finding communities, they calculated for each node $i$ the \textit{within-module degree}, given by the number of its edges that connect to other nodes in the same community, and a \textit{participation ratio} $P_i$, which measures extent to which links are distributed among all network communities.  Guimer\`a and Amaral then interpreted the role of each node based on its location in the plane determined by $P_i$ and the $z$-score $z_i$ of the within-module degree.  They thereby found that non-hub connecting nodes (which have low $z_i$ and moderately high $P_i$, indicating a preferential connectivity to a subset of the network's communities) are systematically more conserved across species than provincial hubs (which have high $z_i$ and low $P_i$).  This appears to be related to the role of non-hub connectors in describing the global structure of fluxes between different network modules.  Importantly, one can follow a similar procedure using other measures, such as betweenness centrality \cite{fenn}, as the essential insight---which, we stress, \textit{was borrowed from ideas in the social sciences}---is to calculate network quantities relative to community assignment.


\subsection*{Legislative Networks} \label{poly}

Advances in network science have also begun to uncover the ways in which
social relationships shape political outcomes \cite{congshort,fowlershort,supreme08}.  In this section, we describe our own work on legislative networks \cite{congshort,conglong,yan,inprep}, in which community detection has played a central role.

Consider a bipartite graph composed of Representatives and their committee and subcommittee (henceforth called simply ``committee") assignments during a single two-year term of the U.~S. House of Representatives. Each edge represents a committee assignment and connects a Representative to a committee.  We project each such network onto a weighted unipartite graph of committees (see Fig.~\ref{intro}), in which the nodes are now committees and the value of each edge gives the normalized connection strength between two committees.  By computing the community structure of these networks and analyzing legislator ideology, we investigated correlations between the political and organizational structure of House committees.  This revealed close ties between the House Rules Committee and the Select Committee on Homeland Security in the 107th (2001-02) and 108th (2003-04) Congresses that broke the established procedures for determining the composition of select committees \cite{congshort,conglong}.  (Figure~\ref{intro} shows the 108th Congress.)  We also showed that the modularity of good network partitions increased following the 1994 elections, in which the Republican party earned majority status in the House for the first time in more than forty years.

Studying networks constructed from legislation cosponsorship can help uncover social connections between politicians, as legislators who work together closely on pieces of legislation are
likely to have friendly (or at least cordial) relations.  Computing centrality measures in these networks gives a who's who list of American politics, as it reveals important players like Bob Dole [R-KA], John McCain [R-AZ], and Ted Kennedy [D-MA] \cite{fowlershort}.  The longitudinal study of community structure in Congressional legislation cosponsorship \cite{yan} and roll-call voting \cite{inprep} networks shows that graph modularity can be used to study partisan polarization and political party realignments.  This reveals patterns suggesting that political parties were not the most significant communities in Congress for certain periods of U.~S. history and that the 1994 party-changing elections \textit{followed} a rise in partisan polarization rather than themselves leading to an abrupt polarization in America.


\section*{Summary and Outlook}

With origins in sociology, computer science, statistics, and other disciplines, the study of network communities is in some respects quite old.  Nevertheless, it has experienced incredible growth since the seminal 2002 paper \cite{structpnas} that brought greater attention to the problem, particularly among statistical physicists \cite{santolong}.  In this survey, we have highlighted an extensive suite of techniques, and there are numerous other methods that we simply haven't had space to discuss (see the review articles \cite{santolong,satu07,commreview} for more information on many of them).  Despite this wealth of technical advances, however, much work remains to be done.  As Mark Newman recently wrote \cite{newmanphystoday}, ``The development of methods for finding communities within networks is a thriving sub-area of the field, with an enormous number of different techniques under development.  Methods for understanding what the communities mean after you find them are, by contrast, still quite primitive, and much needs to be done if we are to gain real knowledge from the output of our computer programs."  One of our primary purposes in writing this article is as a ``call to arms" for the mathematics community to be a part of this exciting endeavor. Accordingly, we close our discussion with additional comments about important unresolved issues.

The remarkable advances of the past few years have been driven largely by a massive influx of data.  Many of the fascinating networks that have been constructed using such data are enormous (with millions of nodes or more).  Given that optimization procedures such as maximizing graph modularity have been proven to be NP-complete \cite{np}, much of the research drive has been to formulate fast methods that still find a reasonable community structure.  Some of the existing algorithms scale well enough to be used on large networks, whereas others must be restricted to smaller ones.  The wealth of data has also led to an increasing prevalence (and, we hope, cognizance) of privacy issues.  However, although the study of network communities has become so prominent, this research area has serious flaws from both theoretical and applied perspectives: There are almost no theorems, and few methods have been developed to use or even validate the communities that we find.

We hope that some of the mathematically-minded \textit{Notices} readers will be sufficiently excited by network community detection to contribute by developing new methods that address important graph features and make existing techniques more rigorous.  When analyzing networks constructed from real-world data, the best practice right now is to use several of the available computationally-tractable algorithms and trust only those structures that are similar across multiple methods in order to be confident that they are properties of the actual data rather than byproducts of the algorithms used to produce them.  Numerous heuristics and analytical arguments are available, but there aren't any theorems, and even the notion of community structure is itself based on the selected methodology used to compute it.  There also appear to be deep but uncharacterized connections between methods that have been developed in different fields \cite{santolong,satu07}.  Additionally, it would be wonderful if there were a clearer understanding of which notions of community and which community-detection methods might be especially appropriate for networks with specific properties and for networks belonging to a specific domain.

At the same time, the problem of how to validate and use communities once they are identified is almost completely open.  Fortunately, recent work offers some hope, as new methods have been developed to infer the existence of missing edges from network data \cite{clausetnature} and relate the composition of communities to intelligent guesses about the demographic characteristics of nodes \cite{facebook}.  (As with social networks more generally, sociologists have already been considering these issues for a long time \cite{butts,freemanbook}.  What we need are techniques that allow us to do this even more effectively.)  In Ref.~\cite{clausetnature}, Aaron Clauset, Cris Moore, and Mark Newman drew on the insight that real-world networks should not be expected to have a unique community structure (despite the output produced by the available methods) and formulated a new \textit{hierarchical random graph model} to develop a method for inferring hierarchical structure from network data.  (A different hierarchical random graph model was formulated for community detection in Ref.~\cite{guimpnas}.)  Their method, which shows excellent promise for future development, allowed them to make accurate predictions about missing connections in partially-known networks.  In our own work on Facebook networks \cite{facebook}, we used permutation tests to further develop methods to quantitatively compare network partitions.  Because one can obtain such partitions either from algorithmic community-detection methods or by hand (from external demographics or other properties), this provides a mechanism to compare the results of different community-finding algorithms and to try to infer node characteristics given partial demographic information.

It is also important to develop community-detection techniques that can be applied to more complicated types of graphs.  As we saw in our discussion of legislative and coauthorship networks,
 collaboration networks have a bipartite structure.  However, there has been seemingly only limited work thus far on community-finding that works directly on bipartite networks rather than on their unipartite projections \cite{barber,guim07,sune,conglong}.  Even fewer community-detection methods are able to handle directed networks (whose adjacency matrices are asymmetric) \cite{guim07,leicht08} or signed networks (whose connections might be ``attracting'' or ``repelling'') \cite{signed}.  Moreover, agents in social networks are typically connected in several different manners---for example, Congressmen can be connected using voting similarities, common committee assignments, common financial contributors, and more---but there are presently very few algorithms that can be applied to such multiplex situations without constructing individual graphs for each category, and further development will likely require the application of ideas from multilinear algebra \cite{kolda,seleekolda}.  It would also be desirable to detect communities in hypergraphs and to be be able to consider connections between agents that are given by interval ranges rather than precise values.  Finally, to be able to study interactions between dynamical processes on networks and the structural dynamics of networks themselves (e.g., if somebody spends a day at home when they have the flu, the network structure in their workplace is different than usual that day), a lot more work is needed on both overlap between communities and on the community structure of time-dependent and parameter-dependent networks.  Analyzing time- and parameter-dependent networks currently relies on \textit{ad hoc} amalgamation of different snapshots rather than on a systematic approach, so it is necessary to develop community-detection methods that incorporate the network structure at multiple times (or parameter values) simultaneously \cite{cosma,vicsek07,hopcroft}.  More generally, this will also have important ramifications for clustering in correlated time series.

We stress that research on network communities has focused on using exclusively structural information (i.e., node connectivity and link weights) to deduce \textit{structural communities} as imperfect proxies for \textit{functional communities} \cite{santolong,facebook,cosma}.  While this seems to be sufficient for some applications \cite{santolong}, in most situations it is not at all clear that structural communities actually map well to the organization of actors in social networks, functions in biological network, etc.  It is hence necessary to develop tools for the detection of functional communities that, whenever possible, incorporate node characteristics and other available information along with the network's structural information.  The elephant in the literature is simply elucidated with just one question: \textit{Now that we have all these ways of detecting communities, what do we do with them?}


\section*{Acknowledgements}

Our views on network community structure have been shaped by numerous discussions with our colleagues and students over the last several years.  We particularly acknowledge Aaron Clauset, Santo Fortunato, Nick Jones, Eric Kelsic, Mark Newman, Stephen Reid, and Chris Wiggins.  We also thank Joe Blitzstein, Tim Elling, Santo Fortunato, James Fowler, A.~J. Friend, Roger Guimer\`{a}, Nick Jones, David Kempe, Franziska Klingner, Renaud Lambiotte, David Lazer, Sune Lehmann, Jim Moody, and David Smith for useful comments on this manuscript, and Christina Frost and Amanda Traud for assistance in preparing some of the figures.  We obtained data from Adam D'Angelo and Facebook, the House of RepresentativesÕ Office of the Clerk (Congressional committee assignments), Mark Newman (network scientist coauthorship), James Fowler (Congressional legislation cosponsorship), and Keith Poole (Congressional roll call votes). PJM was funded by the NSF (DMS-0645369) and by start-up funds provided by the Institute for Advanced Materials, Nanoscience and Technology and the Department of Mathematics at the University of North Carolina at Chapel Hill.  MAP acknowledges a research award (\#220020177) from the James S. McDonnell Foundation.  JPO is supported by the Fulbright Program.





\end{document}